\documentclass[twocolumn,showpacs,showkeys,reprint,amsmath,amssymb,aps]{revtex4-1}

\usepackage{graphicx}
\usepackage{dcolumn}
\usepackage{bm}
\usepackage{tabularx}
\usepackage{amsmath}
\usepackage{xcolor}

\begin{document}

\title{Pressure-induced superconductivity in the van der Waals semiconductor violet phosphorus}

\author{Y. Y. Wu,$^{1,*}$ L. Mu,$^{1,*}$ X. Zhang,$^{1,*}$ D. Z. Dai,$^{1}$ L. Xin,$^{1}$ X. M. Kong,$^{1}$ S. Y. Huang,$^{2,\dag}$ K. Meng,$^{1}$ X. F. Yang,$^{1}$ C. P. Tu,$^{1}$ J. M. Ni,$^{1}$ H. G. Yan,$^{1,3,4}$ and S. Y. Li$^{1,4,5,\ddag}$}

\affiliation
 {$^1$State Key Laboratory of Surface Physics, Department of Physics, Fudan University, Shanghai 200438, China\\
 $^2$Institute of Optoelectronics, Fudan University, Shanghai 200438, China\\
 $^3$Key Laboratory of Micro and Nano-Photonic Structures (Ministry of Education), Fudan Univeristy, Shanghai 200438, China\\
 $^4$Collaborative Innovation Center of Advanced Microstructures, Nanjing 210093, China\\
 $^5$Shanghai Research Center for Quantum Sciences, Shanghai 201315, China}

\date{\today}

\begin{abstract}
Recently, a new transition metal dichalcogenide (TMD) material 2M-phase WS$_2$ has been synthesized. The 2M-WS$_2$ not only exhibits superconductivity, with highest $T${$\rm_c$} of 8.8 K at ambient pressure among TMDs, but also hosts topological surface state. Here we report the low-temperature thermal conductivity measurements on 2M-WS$_2$ single crystals to investigate its superconducting gap structure. A negligible residual linear term $\kappa_0/T$ in zero field shows that 2M-WS$_2$ has fully superconducting gap with all electrons paired. The field dependence of $\kappa_0/T$ suggests anisotropic superconducting gap or multiple nodeless superconducting gaps in 2M-WS$_2$. Such a fully gapped superconducting state is compatible with 2M-WS$_2$ being a topological superconductor candidate.
\end{abstract}

\date{\today}

\begin{abstract}
The van der Waals (vdW) semiconductor black phosphorus has been widely studied, especially after the discovery of phosphorene. On the contrary, its sister compound violet phosphorus, also a vdW semiconductor, has been rarely studied. Here we report the pressure-induced superconductivity in violet phosphorus up to $\sim$40 GPa. The superconductivity emerges at 2.75 GPa, which is well below the structural transition from monoclinic ($M$) to rhombohedral ($R$) structure at 8.5 GPa. The superconducting transition temperature ($T${$\rm_c$}) shows a plateau of $\sim$7 K from 3.6 to 15 GPa, across the $M$ to $R$ structural transition, then jumps to another plateau of $\sim$10 K in the simple cubic ($C$) structure above 15 GPa. The temperature-pressure superconducting phase diagram of violet phosphorus is established, which is different from that of black phosphorus at low pressure. For black phosphorus, the superconductivity emerges until the structural transition from orthorhombic ($O$) to $R$ structure at $\sim$5 GPa, with a lower $T${$\rm_c$} than violet phosphorus. The pressure-induced superconductivity in violet phosphorus demonstrates its tunable electronic properties, and more electronics and optoelectronic applications are expected from this stable vdW semiconductor at ambient conditions.
\end{abstract}

\maketitle

Black phosphorus (BP) is a layered van der Waals (vdW) semiconductor consisting of two-dimensional (2D) atomic layers that are stacked together by weak interlayer coupling \cite{3.2014, 001.2020}. The weak interlayer coupling allows for the exfoliation of layered BP down to monolayer, exhibiting a direct band gap ranging from 0.3 eV in bulk BP to about 2 eV in monolayer BP \cite{00.2014, 89.2014}. The narrow gap of few-layer BP renders it an excellent material for near- and mid-infrared optoelectronics \cite{5.2017}. Moreover, few-layer BP features high carrier mobility \cite{90.2014, 90.2016} and pronounced in-plane anisotropy in optical properties \cite{2.2014, 92.2015, 91.2016}. These characteristics hold potential for the applications of nanophotonics and nanoelectronics \cite{4.2016}. However, the practical applications of few-layer BP are somewhat limited by its instability under ambient conditions \cite{93.2015, 92.2016}.

As a sister compound of BP, another vdW semiconductor violet phosphorus (VP) is the most stable phosphorus allotrope, owing to its higher thermal decomposition temperature than BP \cite{16.2020}. VP with a monoclinic ($M$) layered structure manifests in-plane crystallographic anisotropy and can be be exfoliated to monolayer \cite{16.2020, 27.2022, 90.2022, 3660.2022}. According to the band structure calculation, the bulk VP has an indirect band gap of 1.42 eV, while the monolayer VP has a direct band gap of 2.54 eV and a highly anisotropic hole mobility, making it a good candidate for high frequency electronics and optoelectronic applications in the low wavelength blue color regime \cite{16.2020, 3660.2022, 2975.2016, 91.2022}. The tunable band gap, similar to that of BP, can be controlled by varying the thickness. The photoluminescence (PL) measurements demonstrate an optical band gap of 1.89 eV in 29-layer VP thin flake, slightly larger than 1.77 eV band gap observed in VP single crystal \cite{16.2020, 27.2022}.

Band structure engineering on vdW materials can be achieved through various methods: pressure \cite{007.2000, 008.2018}, intercalation \cite{009.2005, 010.2015}, heterostructure fabrication \cite{011.2015, 012.2014}, and electrical gating \cite{013.2012, 014.2015}. Among them, pressure is a clean one to drive structural transitions and modify the band structure without introducing damages or impurities \cite{51.2020, 52.2017}. For instance, BP has an orthorhombic ($O$) structure at ambient pressure. Under high pressure, BP undergoes a structural transition to rhombohedral ($R$) structure at 5 GPa, and then to simple cubic ($C$) structure at 12 GPa \cite{23.1963, 6.1984, 28.1985, 29.2002, 10.2017}. At 1.2 GPa, there is a transition from semiconductor to metal \cite{9.2015, 12.2016, 11.2018}. Superconductivity appears at the critical pressure 5 GPa of the $O$-$R$ structural transition, with a superconducting critical temperature ($T${$\rm_c$}) of 3.2 K \cite{10.2017}. As pressure further increases, $T${$\rm_c$} gradually increases and finally saturates at $\sim$10 K above 30 GPa \cite{6.1984, 28.1985, 29.2002, 10.2017, 39.2017}. For VP under pressure, so far there is only a Raman scattering study, which shows two pressure-induced structural transitions: the $M$ structure transforms into the $R$ structure at about 8.5 GPa, then to the $C$ structure at pressure above 13.6 GPa \cite{17.2021}. The study on the electrical transport properties of VP under pressure is still lacking.

In this Letter, we perform the high-pressure measurements of electrical resistance and PL spectra on VP, and construct its superconducting phase diagram under pressure. Superconductivity with $T${$\rm_c$} of 4.3 K is found to emerge at 2.75 GPa in the $M$ structure of VP. Then the $T${$\rm_c$} shows two plateaus: one around 7 K across the $M$-$R$ structural transition, and the other around 10 K in the $C$ structure. We compare it with the superconducting phase diagram of BP under pressure.

\begin{figure}
\includegraphics[clip,width=8cm]{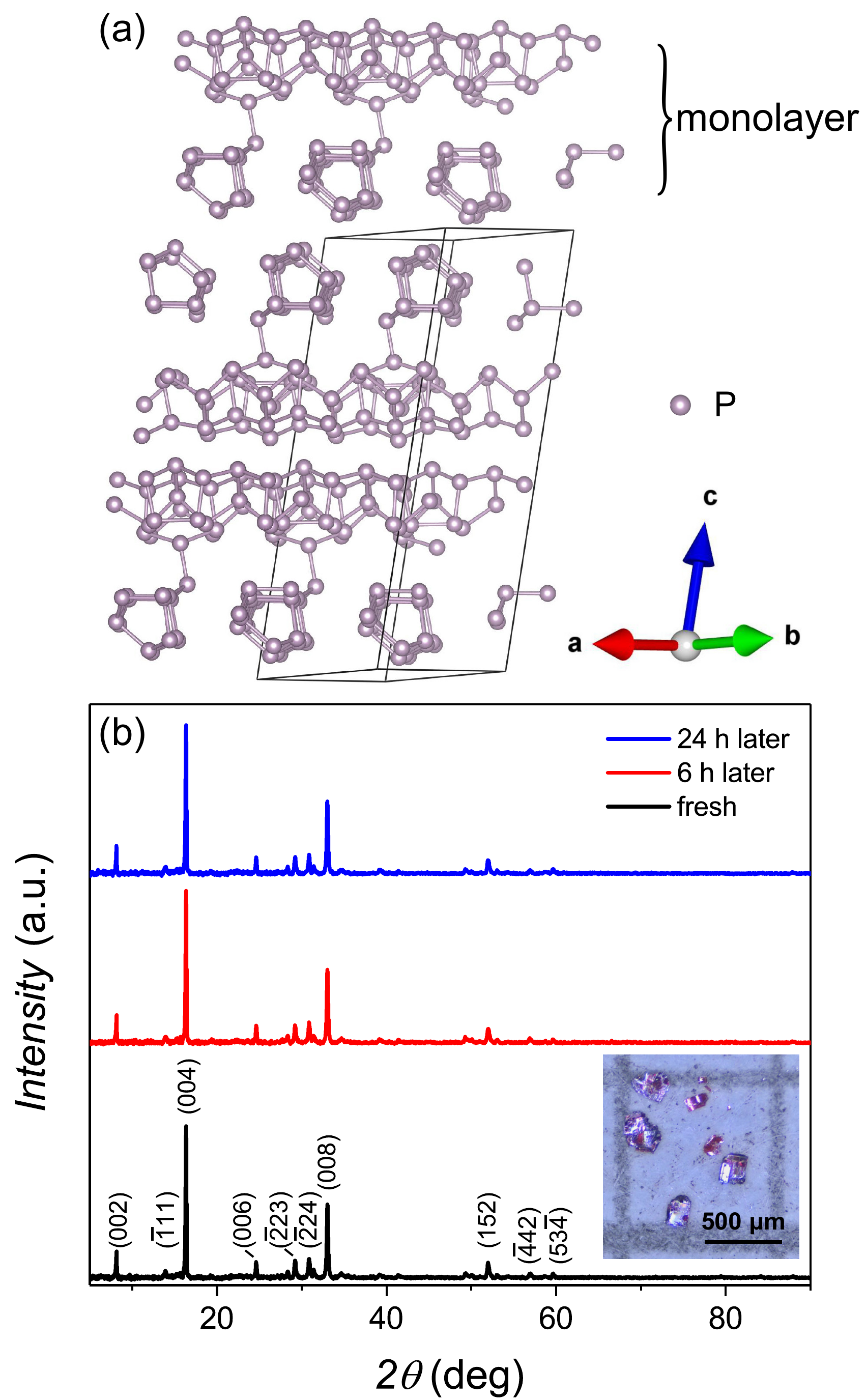}
\caption{(a) The monoclinic crystal structure of violet phosphorus. It manifests layered structure and the monolayer is clearly shown. The box depicts the monoclinic unit cell. (b) X-ray diffraction patterns of violet phosphorus powder, obtained by crushing single crystals. The black, red, and blue lines represent the fresh sample, exposed to air for 6 and 24 hours, respectively. No impure phases and no change over time are found from these patterns, demonstrating that violet phosphorus is stable in air for at lease one day. Inset is a photograph of the violet phosphorus single crystals.}
\end{figure}

VP single crystals were synthesized by the chemical vapor transport (CVT) method \cite{16.2020}, using red phosphorus as the source and Sn$+$SnI$_4$ as transport agents. A mixture of red phosphorus (470 mg, Aladdin, 99.999$\%$), Sn (10 mg, Alfa Aesar, 99.995$\%$), and SnI$_4$ (18 mg, Alfa Aesar, 99.998$\%$) was sealed in an evacuated quartz tube. The quartz tube was slowly heated up for 8 h until one end of the quartz tube with phosphorus source to 650 $^\circ{\rm C}$ and the other empty end to 630 $^\circ{\rm C}$. The quartz tube was kept at these temperatures for 5 h and then slowly cooled 100 $^\circ{\rm C}$ down during 10 h. The end with phosphorus source was then kept at 550 $^\circ{\rm C}$ with the other end at 530 $^\circ{\rm C}$ for 30 h. The quartz tube was then slowly cooled to room temperature during 70 h. Both VP and BP were obtained. The yield of VP is about 36\% and was then separated under a microscope.

The X-ray diffraction (XRD) measurement was performed on an X-ray diffractometer (Bruker D8 Advance) with Cu K$\alpha$ radiation (wavelength $\lambda$ = 1.542 \AA). High-pressure electrical resistance was measured in a Cu-Be alloy diamond anvil cell (DAC) with a 300 $\mu$m culet size. The Be-Cu plate was used as gasket and powdered cubic boron nitride served as insulating material. Small piece of VP single crystal was loaded inside the DAC and no pressure transmitting medium was used. The sample was crushed into powder in DAC upon compression. Platinum foils with a thickness of 4 $\mu$m were used for the electrodes. The standard van der Pauw method was adopted to measure the resistance by four-probe configuration. Low-temperature resistance measurement was carried out in a physical property measurement system (PPMS, Quantum Design). High-pressure PL spectrum was detected in a stainless-steel DAC with a 500 $\mu$m culet size. The selected VP flake was transferred to the diamond culet by PDMS dry transfer method. The stainless-steel plate was used as gasket and silicone oil was used as an inert pressure transmitting medium. Room-temperature PL spectrum was detected by a confocal Raman microscope (Horiba Jobin-Yvon Labram HR Evolution). The 532 nm laser was available for excitation. A ruby ball was used to calibrate the pressure at room temperature before and after the experiments. The thickness of VP flake on SiO$_2$/Si substrate was measured by an atomic force microscope (AFM, Park NX10) in tapping mode and at ambient conditions.

Monoclinic VP belongs to space-group ${P2/n}$ (No. 13) and has a more complex atomic arrangement compared to BP \cite{16.2020, 27.2022}. The layered structure of VP is sketched in Fig. 1(a), which also clearly displays one monolayer \cite{21.1969}. Due to the small size of the single crystals (the inset of Fig. 1(b)), obtaining a high-resolution XRD pattern of the natural or cleaved (00$l$) plane is challenging. Therefore, some small pieces of VP single crystals were crushed into powder for XRD measurement. Figure 1(b) displays the XRD patterns during different time of exposure to air for the VP powder. One can see that there is no impure phase and VP is stable in air for at lease one day. Nevertheless, we completed the DAC assembly within 6 h to ensure the high purity of VP sample.

\begin{figure}
\includegraphics[width=8.5cm]{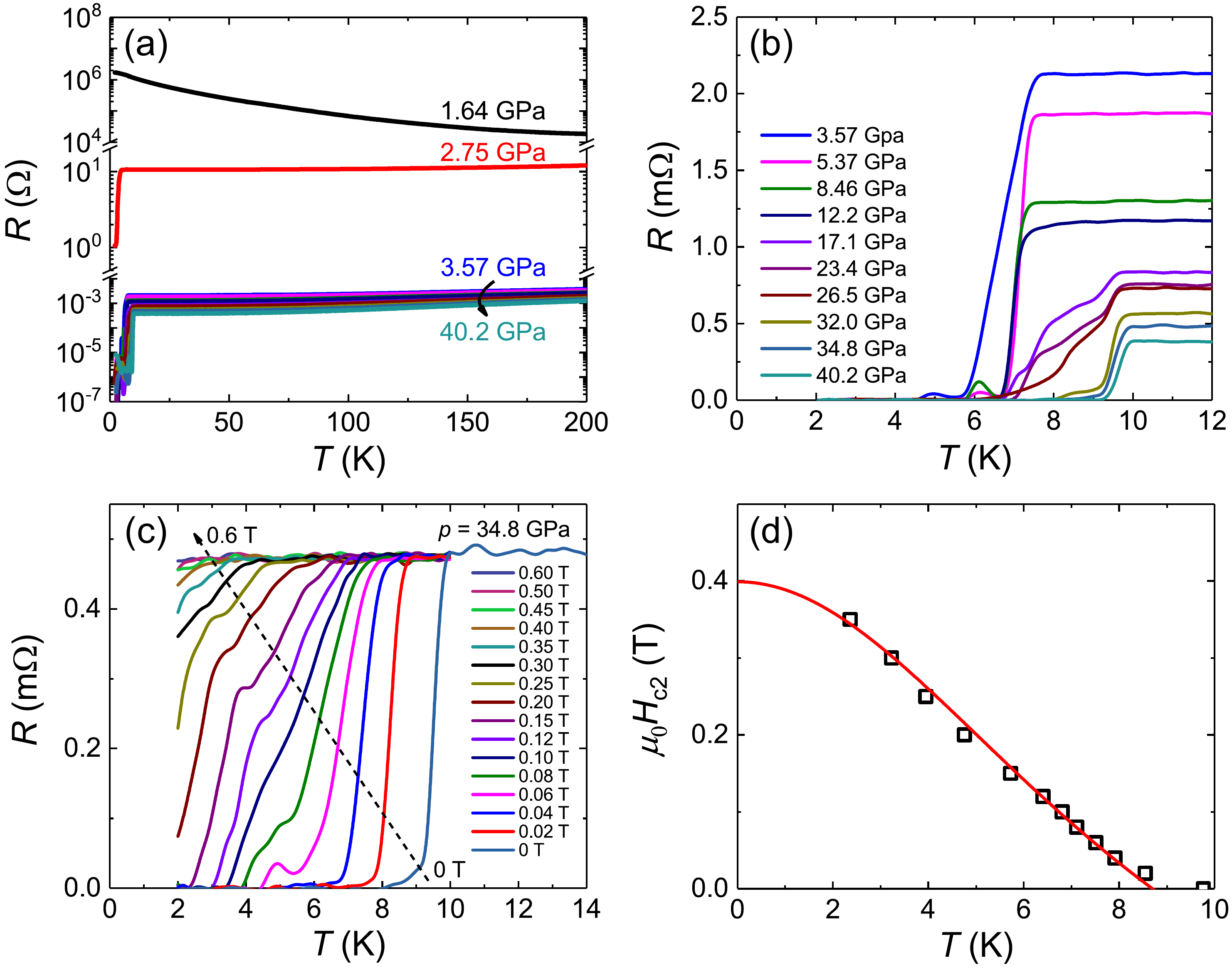}
\caption{(a) Temperature dependence of resistance for violet phosphorus under pressures up to 40.2 GPa. (b) The low temperature superconducting transitions for pressures between 3.57 and 40.2 GPa. (c) Temperature dependence of resistance in different magnetic fields under the pressure of 34.8 GPa. The superconducting transition is gradually suppressed to lower temperature with increasing the field. (d) Temperature dependence of the upper critical field $\mu_0H{\rm_{c2}}$ under 34.8 GPa. The superconducting transition temperature ($T${$\rm_c$}) is defined at the 10$\%$ drop of the normal-state resistance ($T${$\rm_c$}$^{10\%}$). The red line is a fit to the Ginzburg-Landau formula, $\mu_0H{\rm_{c2}}(T)=\mu_0H{\rm_{c2}}(0)(1-(T/T{\rm_c})^2)/(1+(T/T{\rm_c})^2)$.}
\end{figure}

\begin{figure}
\includegraphics[width=6.1cm]{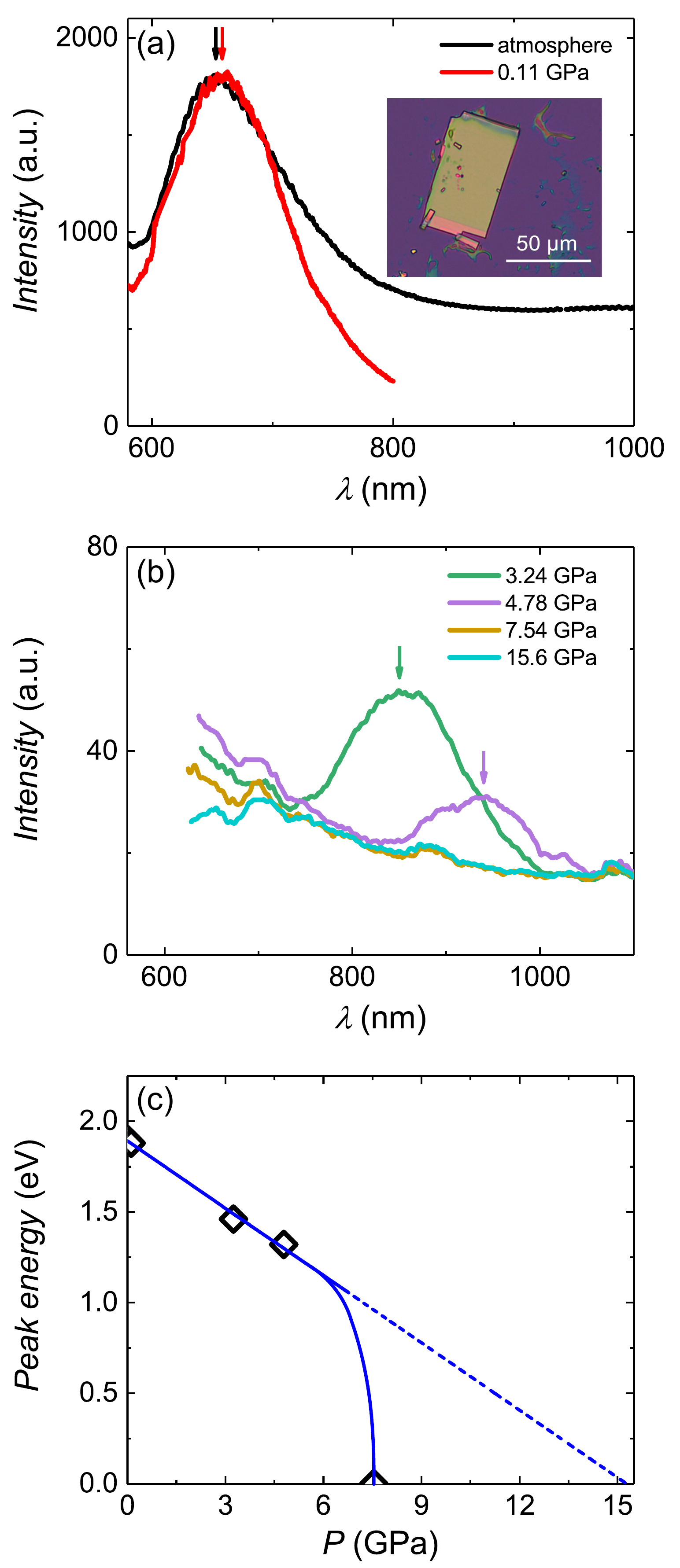}
\caption{(a) and (b) Photoluminescence (PL) spectra of violet phosphorus flake under various pressures at room temperature. The PL emissions from ruby and diamond substrate were subtracted and the arrow indicates the maximum intensity of PL peak, which reflects the optical band gap at low pressures. The small pressure 0.11 GPa is generated during assembling the diamond anvil cell. Inset is a photograph of violet phosphorus flake on SiO$_2$/Si substrate with a thickness of 188 nm (99 layers). (c) Pressure-dependent peak energy of violet phosphorus flake. The peak energy shows a linear dependence at low pressures and the coefficient is $-$0.124 eV/GPa. At 7.54 GPa, the optical band gap disappears, and the curved solid line is a guide for the eye.
}
\end{figure}

Figure 2(a) shows the temperature dependence of resistance $R$($T$) for VP under pressures up to 40.2 GPa. At 1.64 GPa, the $R$($T$) curve still manifests an insulating behavior. When pressure is increased to 2.75 GPa, a large resistance drop at 4.3 K is observed, indicating the emergence of superconductivity. With further increasing pressure, zero resistance is obtained at low temperatre. In Fig. 2(b), the low-temperature superconducting transitions are plotted for pressures between 3.57 and 40.2 GPa. As the pressure increases, the normal-state resistance deceases monotonically, and the transition temperature $T${$\rm_c$} increases from 7.3 to 9.8 K. Here, $T${$\rm_c$} is defined as the 10$\%$ drop of the normal-state resistance ($T${$\rm_c$}$^{10\%}$). At 17.1 GPa, the transition becomes broad and shows steps. This usually implies the coexistence of different superconducting phases \cite{22.2021, 23.2011, 24.2017}, likely owing to the $R$-$C$ structural transition of VP and some degree of pressure inhomogeneity. Above 32.0 GPa, this broadening and the step transition disappear, with $T${$\rm_c$} slightly shifting to 9.8 K. From 32.0 to 40.2 GPa, $T${$\rm_c$} almost saturates in the $C$ structure.

The low-temperature superconducting transitions in different magnetic fields under the pressure of 34.8 GPa are plotted in Fig. 2(c). The transition is gradually suppressed to lower temperature with increasing the field and it completely vanishes in 0.50 T. The temperature dependence of the upper critical field ($\mu_0H{\rm_{c2}}$) at 34.8 GPa is summarized in Fig. 2(d). The data can be well fitted by the empirical Ginzburg-Landau (GL) formula $\mu_0H{\rm_{c2}}(T)=\mu_0H{\rm_{c2}}(0)(1-(T/T{\rm_c})^2)/(1+(T/T{\rm_c})^2)$ and the $\mu_0H{\rm_{c2}}(0)$ is determined to be 0.40 T.

The superconducting transition is occasionally accompanied by a structural transition. In the case of BP, the superconductivity is observed at the critical pressure of $O$-$R$ structural transition \cite{10.2017}. For VP, it remains the $M$ structure until 7.44 GPa and a phase transition from semiconducting $M$ structure to metallic $R$ structure occurs between 7.44 - 8.49 GPa, as demonstrated by the pressure-dependent Raman spectroscopy \cite{17.2021}. Different from BP, we find that the superconducting transition in VP occurs at 2.75 GPa, well below the critical pressure of $M$-$R$ structural transition.

To confirm the order of superconducting and structural phase transitions in VP, PL spectra under various pressures are measured, as shown in Figs. 3(a) and 3(b). PL spectroscopy is an ubiquitous tool used to study exciton emission in semiconductor and the emission energy is usually termed as optical band gap. Hence, PL spectroscopy is widely employed to probe the band structures \cite{27.2022, 26.2010}. As seen in Figs. 3(a) and 3(b), a prominent PL peak for VP flake (99 layers) is detected with the maximum intensity at $\sim$653 nm (1.9 eV) under ambient pressure, which is consistent with previously reported optical band gap in 29-layer VP thin flake \cite{27.2022}. The detected peak energy is corresponding to the indirect band gap of VP. The assembly of DAC generates a small pressure of 0.11 GPa, making the PL peak slightly red-shifted. Upon further compression, the PL peak shifts to 850 (1.46 eV) and 940 nm (1.32 eV) under 3.24 and 4.78 GPa, respectively. The pressure dependence of the peak energy is given in Fig. 3(c). The peak energy shows a linear dependence at low pressures and the coefficient is estimated to be $-$0.124 eV/GPa, which is smaller than that of BP \cite{20.2021, 99.2001}. As shown in Fig. 3(b), the optical band gap in VP exists until 4.78 GPa, suggesting that the phase transition from semiconducting $M$ structure to metallic $R$ structure has not occurred under this pressure. Therefore, we confirm that VP exhibits superconductivity in the well-defined $M$ structure at 2.75 GPa.

\begin{figure}
\includegraphics[width=8.7cm]{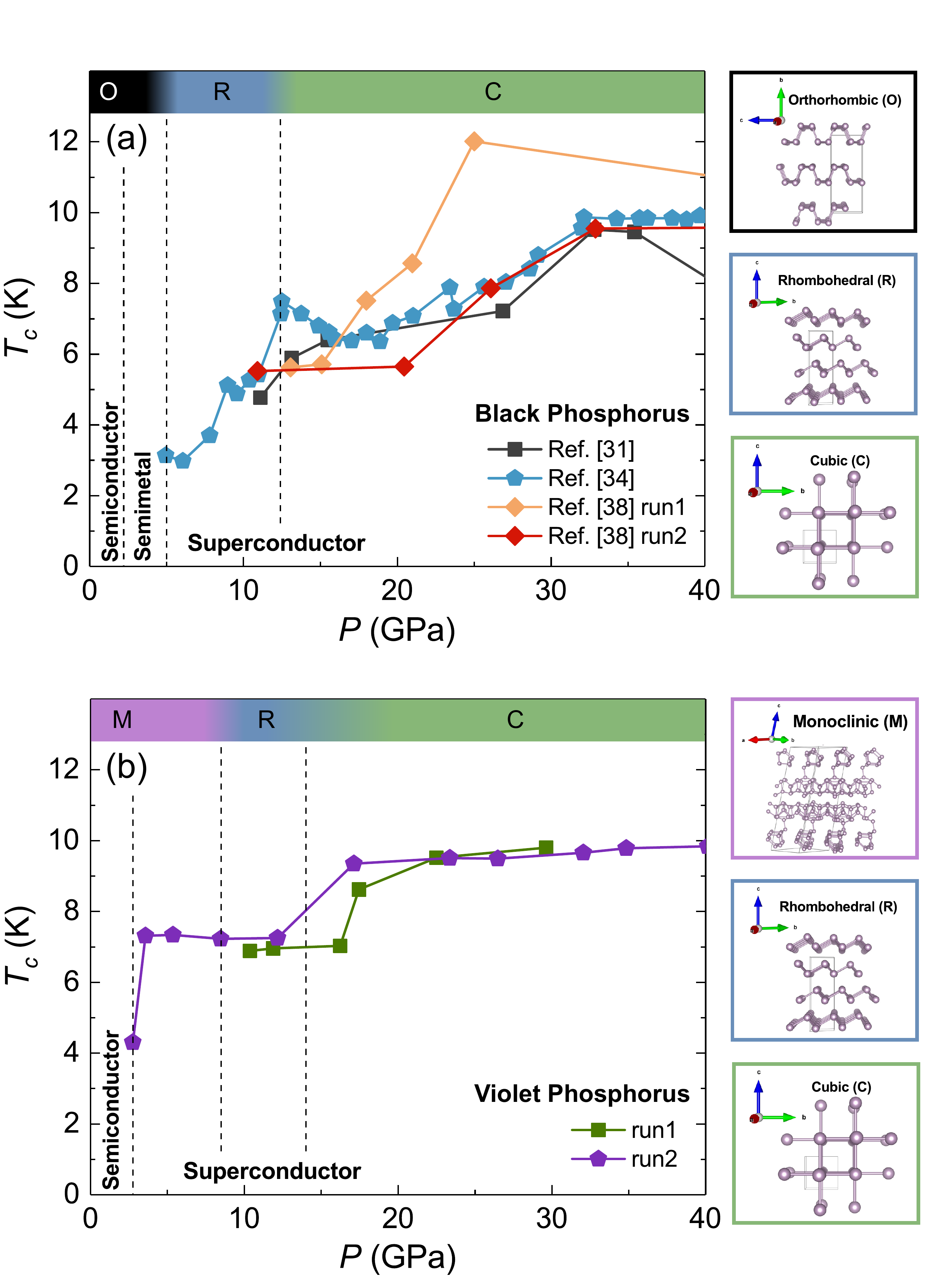}
\caption{Phase diagrams of (a) black phosphorus and (b) violet phosphorus under pressure. $O$, $M$, $R$, and $C$ represent orthorhombic, monoclinic, rhombohedral and simple cubic structures, as shown on the right of the panels. Under pressure, violet phosphorus manifests superconductivity in the monoclinic structure, while black phosphorus becomes superconducting until the transition from orthorhombic to rhombohedral structure. The data of black phosphorus are from Refs. \cite{29.2002, 10.2017, 39.2017}, and the pressure evolution of structure for violet phosphorus is from Ref. \cite{17.2021}.}
\end{figure}

While both VP and BP are forms of phosphorus and exhibit superconductivity under pressure, they have different superconducting phase diagrams under pressure, as plotted in Fig. 4. BP has three discernible states under pressure from the electronic behavior: semiconducting, semimetallic and superconducting states (Fig. 4(a)). At 1.2 GPa, BP undergoes a transition from semiconducting to semimetallic state \cite{9.2015, 12.2016, 11.2018}. Superconductivity occurs at 5 GPa with a $T${$\rm_c$} of 3.2 K, coinciding with the $O$-$R$ structural transition \cite{23.1963, 6.1984, 28.1985, 29.2002, 10.2017}. Then the $T${$\rm_c$} gradually increases to about 10 K before finally saturating above 30 GPa \cite{6.1984, 28.1985, 29.2002, 10.2017, 39.2017}. In contrast, VP manifests semiconducting and superconducting states upon compression (Fig. 4(b)). Superconductivity appears at 2.75 GPa in the $M$ structure, with a $T${$\rm_c$} of 4.3 K. $T${$\rm_c$} then increases to 7.3 K and stays constant over a broad pressure range from 3.57 to 12.2 GPa, during which VP undergoes the $M$-$R$ structural transition at 8.5 GPa. At 17.1 GPa, $T${$\rm_c$} jumps to 9.4 K in the $C$ structure. Thus VP has a higher $T${$\rm_c$} than BP at low pressure, and both of them show a saturated $T${$\rm_c$} at $\sim$10 K at high pressure above 30 GPa.

In summary, we have performed high-pressure resistance measurements on the vdW semiconductor VP up to 40.2 GPa. It is found that superconductivity with $T${$\rm_c$} $=$ 4.3 K emerges at a pressure of 2.75 GPa, well before the $M$-$R$ structure transition. This is confirmed by our PL spectroscopy measurements. From 3.57 to 12.2 GPa, $T${$\rm_c$} stays at a plateau of $\sim$7 K across the $M$-$R$ structural transition. Then, $T${$\rm_c$} increased to another plateau of $\sim$10 K in the $C$ structure at 17.1 GPa, and remains this saturation up to 40.2 GPa. Sucn an superconducting phase diagrma under pressure is different from its sister compound BP, which deserves further theoretical and experimental studies to clarify the origin.

This work was supported by the Natural Science Foundation of China (Grant No. 12174064), the National Key Research and Development Program of China (Grant No. 2021YFA1400100), and the Shanghai Municipal Science and Technology Major Project (Grant No. 2019SHZDZX01). Part of the experimental work has been carried out in Fudan Nanofabrication Laboratory. We thank Nanjing 2DNANO Tech. Co., Ltd. for providing us violet phosphorus single crystals.

\noindent $^*$ These authors contributed equally to this work.

\noindent $^\dag$ E-mail: syhuang@fudan.edu.cn

\noindent $^\ddag$ E-mail: shiyan$\_$li$@$fudan.edu.cn

\end{document}